D.M. Zayachuk[1], V.E. Slynko[2], and A. Csik[3]

# Morphology of PbTe Crystal Surface Sputtered by Argon Plasma under Secondary Neutral Mass Spectrometry Conditions

[1]*Lviv Polytechnic National University, Lviv, Ukraine*
[2]*Institute of Material Science NASU, Chernivtsy, Ukraine*
[3]*Institute for Nuclear Research, Hungarian Academy of Sciences (ATOMKI), Debrecen, Hungary*

We have investigated morphology of the lateral surfaces of PbTe crystal samples grown from melt by the Bridgman method sputtered by $Ar^+$ plasma with ion energy of 50 – 550 eV for 5 - 50 minutes under Secondary Neutral Mass Spectrometry (SNMS) conditions. The sputtered PbTe crystal surface was found to be simultaneously both the source of sputtered material and the efficient substrate for re-deposition of the sputtered material during the depth profiling. During sputtering PbTe crystal surface is forming the dimple relief. To be re-deposited the sputtered Pb and Te form arrays of the microscopic surface structures in the shapes of hillocks, pyramids, cones and others on the PbTe crystal sputtered surface. Correlation between the density of re-deposited microscopic surface structures, their shape, and average size, on the one hand, and the energy and duration of sputtering, on the other, is revealed.

## Introduction

SNMS is known to be an effective method for composition analysis and element depth profiling of multicomponent and doped solids [1-3]. Its high sensitivity (down to a ppm range) makes it suitable for successful analysis of solid surfaces and thin films, for the elemental characterization of samples in different fields [4-8]. One of the preconditions for the successful application of SNMS method for the correct quantitative composition analysis is having the flat surface to be sputtered by plasma ions. However, for many practical objectives this requirement cannot be accomplished a priori. In particular, such a situation occurs in the case of lateral profiling of crystal ingots having a conical-cylindrical shape. The need for such investigations arises, primarily, in studying the distribution of doping impurities in the surface layers of crystals grown from doped melts. In particular, such studies recently succeeded in establishing the unique behavior of the rare earth element Eu impurity introduced in small (at the level of $10^{19}$ $cm^{-3}$) amounts into the initial melt for growth of PbTe:Eu doped crystals by the Bridgman method [9].

Profiling to a large depth (in tens of micrometers) requires prolonged surface sputtering – tens of minutes or more. Even if the initial surface of the sample is flat, there is no guarantee that it will remain such throughout the sputtering process. This can be especially relevant in the studies of multicomponent systems with preferential sputtering [1, 10, 11]. In particular, when PbTe crystals are sputtered under SNMS conditions, extremely nontrivial features of formation of Pb and Te sputtered phase arise such as the huge preference of Te sputtering under low sputtering energy below 160 eV, oscillations of Pb and Te sputter yield during prolonged sputtering, decrease of the average intensity of Pb and Te signals with sputtering time at low sputtering energy and increase at high [12]. One of the reasons for such behaviour of sputtered phase of PbTe crystals is the state of their surface, which is covered by the array of crystalline formations re-depositing on the sputtering surface during the depth profiling experiment. In this work we continue our investigation of sputtering of PbTe crystals by $Ar^+$ ions in the SNMS conditions started in [12]; it is devoted to detailed study of morphology of the sputtered lateral surfaces of the crystals grown from melt by the Bridgman method.

## I. Experiment

The PbTe crystal ingot used for investigation was grown from melt by the Bridgman method. High-purity lead and tellurium were used for PbTe crystal ingot growing. Despite a rather low content of background impurities additional cleaning of the initial materials was applied. Special attention was given to its clearing from oxygen, because both lead and tellurium easily oxidized in air. The quartz ampoule, as usual coated inside with a thin layer of pyrolytic graphite, has been used for growing crystal.

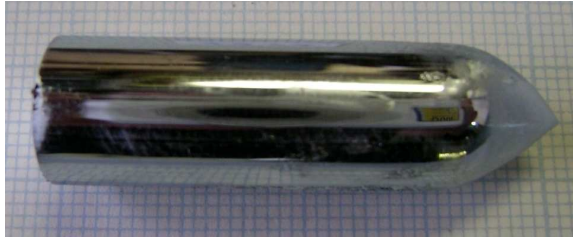

**Fig. 1.** The general view of the PbTe crystal ingot, which was used for manufacturing of the samples for investigations.

The general view of the grown ingot is shown in Fig. 1.

The ingot had a conical-cylindrical shape. The diameter of its cylindrical part using for manufacturing of the samples for investigations was 11.5 mm. Such transverse dimensions of the crystal ingot were sufficient to provide quite suitable conditions for sputtering of the lateral crystal surface by the beam of $Ar^+$ ions through a Ta mask with an aperture of diameter of 2 mm. On the other hand, cylindrical shape of the crystal ingot gave an opportunity to carry out sputtering of PbTe crystal samples along different crystallographic directions, thus examining impact of the sputtering direction on the result of sputtering of the PbTe crystal surface.

The lateral surfaces were sputtered. Sputtering experiments were carried out on INA-X type SNMS system produced by SPECS GmbH, Berlin. The surface bombardment and post-ionization of sputtered neutral particles were done at low pressure by Electron Cyclotron Wave Resonance (ECWR) argon plasma. In the direct bombardment mode, $Ar^+$ ions are extracted from low pressure plasma and bombard a negatively biased (-50…-550 V) sample surface with a current density of ~ 1 mA/cm$^2$, performing a controlled surface erosion. The sputtered area was confined to a circle of 2 mm in diameter by a Ta mask. Post-ionized neutral particles are directed into a quadruple mass spectrometer Balzers QMA 410 by electrostatic lenses and a broad-pass energy analyzer.

The surface morphology of samples after ion sputtering has been analyzed by Scanning Electron Microscopy (SEM, Hitachi S-4300 CFE). The composition of the samples was verified by Energy Dispersive X-ray (EDX) analysis.

## II. Results

Systematic SEM studies of the surfaces of PbTe crystal samples sputtered by $Ar^+$ plasma at different energies and time showed that the surface morphology significantly changes under impact of $Ar^+$ beams during the sputtering process.

The initial lateral surfaces of the samples had a characteristic metallic gloss; they were smooth and contained some number of growth defects caused as usual by contact with the growth ampoule and the crystal surface. The same surfaces sputtered by $Ar^+$ beam always had a characteristic dimple relief and were covered by the arrays of small surface structures, like hillocks, pyramids, cones and others. Fig. 2 shows some examples for SEM images of the crystal surfaces after their sputtering by Ar plasma with 350 eV ions energy throughout the different sputtering times from 5 till 50 minutes. Fig. 3 shows the impact of different sputtering energy on the state of lateral PbTe crystal surfaces sputtered for the same time, namely 50 minutes.

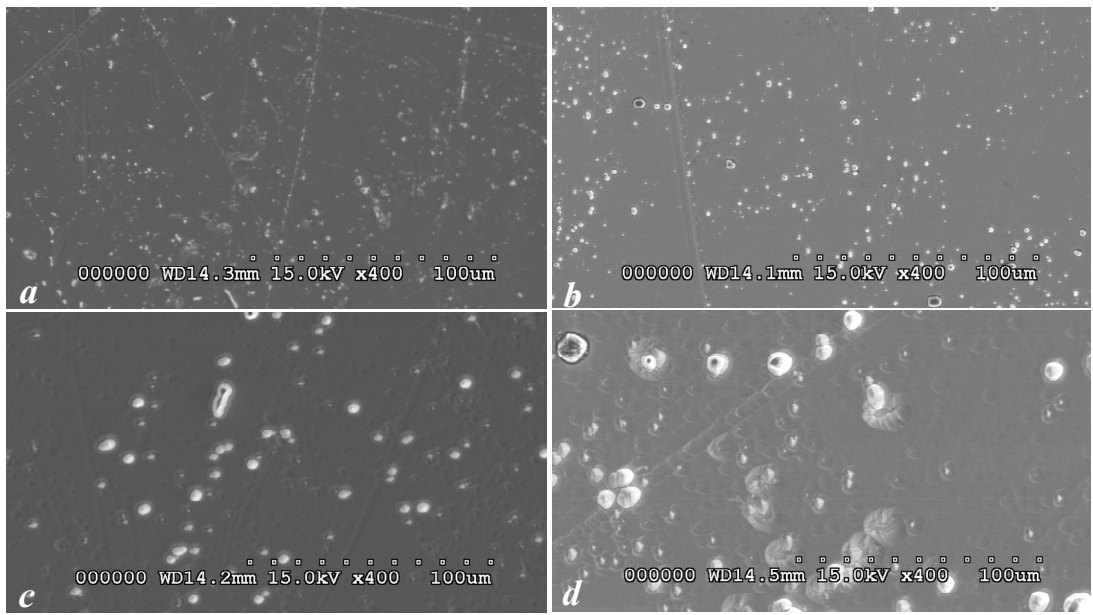

**Fig. 2.** SEM images of lateral surface of PbTe crystal after sputtering by Ar plasma with 350 eV ions energy during: *a* – 5 min; *b* – 15 min; *c* – 35 min; *d* – 50 min.

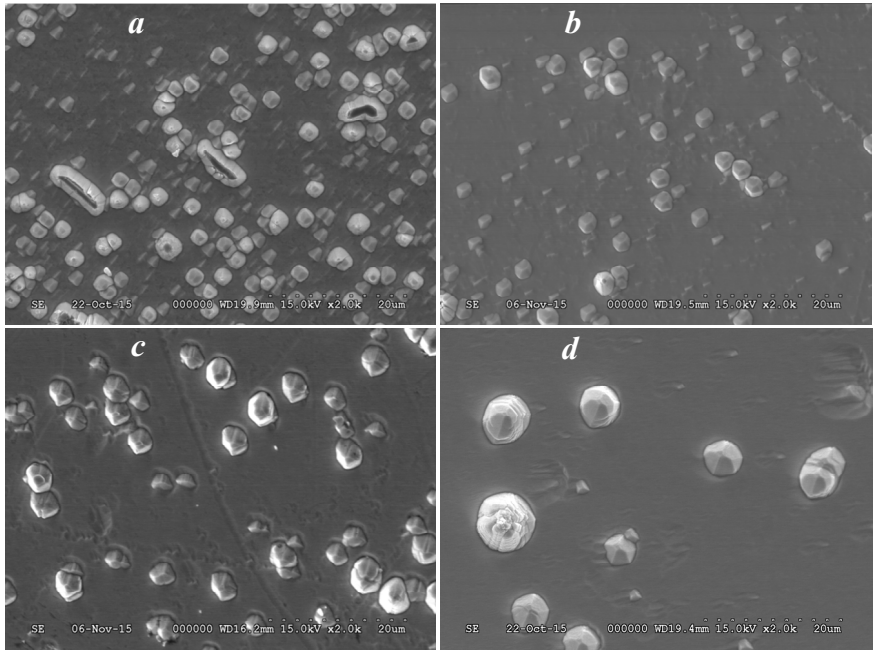

**Fig. 3.** SEM images of the lateral surface of PbTe crystal after sputtering by Ar plasma for 50 min, with ion energy: *a* – 50 eV; *b* – 80 eV; *c* – 160 eV; *d* – 550 eV.

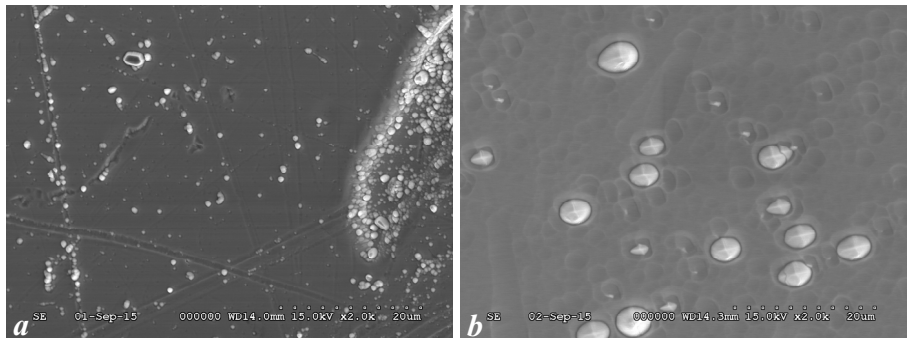

**Fig 4.** SEM images of the lateral surface of PbTe crystal after sputtering by Ar plasma with 350 eV ions energy during: *a* – 5 min; *b* – 25 min.

One can see that the relief of sputtered surfaces as well as the average size, shape, and density of the surface structures re-deposited on the sputtered surface are significantly modified when both duration of sputtering and sputtering energy are changing.

The main experimental results can be systematized as follows.

**2.1. Duration of sputtering and relief of the sputtered surfaces.**

The minimum sputtering time used in our experiments was five minutes. During this time the relief of sputtering surface formed by the $Ar^+$ ions consists in array of incipient dimples with low surface density (Fig. 2, *a*). In the next experiment duration of sputtering was 15 minutes. With such sputtering duration the surface dimples become significantly larger in both the area and the depth. Their surface density increases too, but the dimples are still for the most part spatially separated from each other (Fig 2,*b*). At longer sputtering (35 minutes and more), practically the entire crystal surface covered with the dimples of various area and depth, which are overlapped with each other (Fig 2, *c*, *d*).

**2.2. Sputtering and re-deposition of the sputtered components**

Simultaneously with forming of the surface relief under impact of $Ar^+$ plasma, re-deposition of the sputtered material onto the sputtering surface is taking place. In this study we will focus only on those aspects of this phenomenon which have direct relation, in our opinion, to the peculiarities of the formation of Pb and Te sputtered phase in the SNMS method in quantitative analysis of PbTe samples, which we reported in [12].

For any duration of sputtering of PbTe crystal during the depth profiling measurement, its resulting surface is always covered with some array of microscopic re-deposition structures formed on the surface in the process of its sputtering. When duration of surface sputtering is smaller, the average size of re-deposited microscopic structures is smaller as well, and their surface density is higher (Fig. 2). The re-deposited structures are localized on the natural growth defects of crystal surface as well as within the limits of the dimples of relief of sputtered

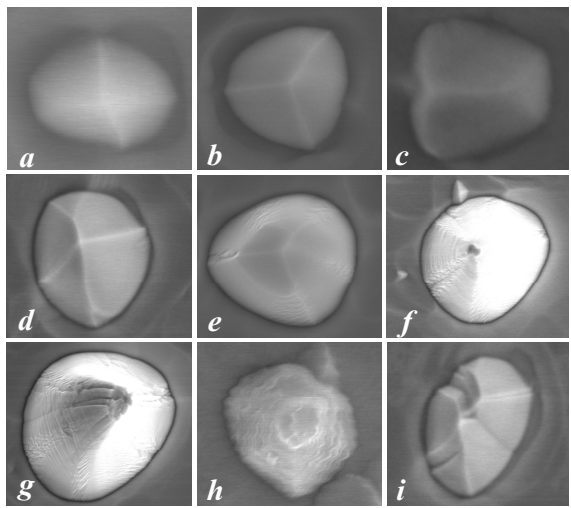

**Fig. 5.** SEM images of the predominant modifications of the surface structures forming on the dimple relief of PbTe sputtered surface.

lateral surface (Fig. 4).

The re-deposited surface structures associated with the natural growth defects of the initial crystal surface have often no certain dominant forms and spatial orientation. The view of re-deposited surface structures associated with the dimple relief of sputtered surface depends first of all on the duration of both sputtering and re-deposition as well as the crystallographic orientation of the surface serving as a substrate.

For the short sputtering and re-deposition time of about 5 minutes, it is difficult to discern any particular faceting and orientation of the re-deposited structures. If the sputtering time is 15 minutes they already become quite strongly pronounced. The dominant forms are pyramidal structures: the quadrangular pyramids corresponding to the growth directions <100> (Fig. 5, *a*); the triangular pyramids corresponding to the growth directions <111> (Fig 5, *b*); or the pyramids put on edge corresponding to the growth directions <110> (Fig 5, *c*) for the crystals with cubic symmetry. After the prolonged sputtering of the sample, accompanied by the increase of the sizes of re-deposited surface structures, a considerable variety of the structure shapes is observed – additional faceting of the structures (Fig. 5, *d*); cone-shaped structures with a pyramidal (Fig 5*e*), conical (Fig 5*f*) or elongated (Fig 5, *g*) tops; pyramids of hexagonal type (Fig 5, *h*); coalescence without re-crystallization of the separate surface structures in larger formations (Fig. 5, *i*).

At constant duration of sputtering of PbTe crystal lateral surfaces the final surface density of re-deposited structures formed on the sputtering surface, shape of the structures, their average size and size distribution depend essentially on the sputtering energy (Fig. 3). Comparing the SEM images of the PbTe crystal surfaces after their sputtering by Ar$^+$ ions with energies from 50 to 550 eV throughout the 50 minutes of sputtering times (Fig. 3), it can be seen that when the sputtering energy is lower, the density of surface structures re-deposited on the sputtered surface is higher. It can be also found that the average sizes of the surface structures are smaller. At the same time the probability of the simultaneous existence of the surface structures of different sizes is higher.

The size range of the PbTe surface structures re-deposited on the lateral crystal surface during its sputtering is very wide on any stage of sputtering process for any sputtering energy (Fig. 2 - 4).

## III. Discussion

Formation of surface reliefs, hillocks, pyramids, cones, whiskers and other structures on the sputtered surface of solids under impact of ions is a common phenomenon [13-17]. Intense re-deposition of the sputtered Pb and Te on the sputtering PbTe crystal surface shows that the sputtered phase over the sputtering surface is strongly supersaturated. A specific feature of this re-deposition is that that it takes place in the conditions of continuous sputtering of re-depositing materials. Let us analyze the obtained results based on these two obvious inferences.

As is well known, deposition of any structures on solid substrate starts from nucleation of depositing material. High density of the structures re-deposited on sputtering surface at the initial stages of sputtering indicates that on the initial lateral surface of PbTe crystal grown from melt by the Bridgman method there are a lot of effective nucleation centers. Decrease of density of the re-deposited microscopic structures accompanied by increase the structure sizes under impact of the stable Ar$^+$ plasma during prolonged crystal depth profiling suggest that, firstly, the vast majority of the nuclei of re-deposited structures are formed on the sputtering lateral surface of PbTe crystal at the early stages of re-deposition, and, secondly, the longer the PbTe crystal surface is sputtered, the less efficiently the new nuclei of re-depositing phase are formed on it.

We believe that decrease of effectiveness of relief dimples as the nucleation centers for re-deposition of sputtered Pb and Te during prolonged PbTe crystal depth profiling is due to the features of both the phase diagrams of PbTe and the PbTe crystals sputtering. PbTe is a semiconductor with a quite narrow homogeneity range [18]. Therefore, composition of PbTe crystals grown by any technological method is always close to stoichiometric [19]. This should also take place in the process of re-deposition of Pb and Te from the sputtered phase. To form a stoichiometric nucleus, Pb and Te atoms should be re-deposited on the sputtering surface in equal amounts. As we have reported recently [12] sputtering of the PbTe crystal is preferential. Preferential sputtering changes the ratio between component concentrations on the sputtering surface of multicomponent material, which becomes more pronounced during longer sputtering. In the initial state the surface concentrations of Pb and Te in the PbTe crystal are equal to each other. Therefore the little dimple on such surface is the best place for nucleation of PbTe re-deposited phase. For sputtered PbTe crystal surface the Pb and Te concentrations differ from each other, and the longer sputtering, the stronger. So, during the prolonged sputtering, nucleation on the sputtered PbTe crystal surface under the same level of supersaturation

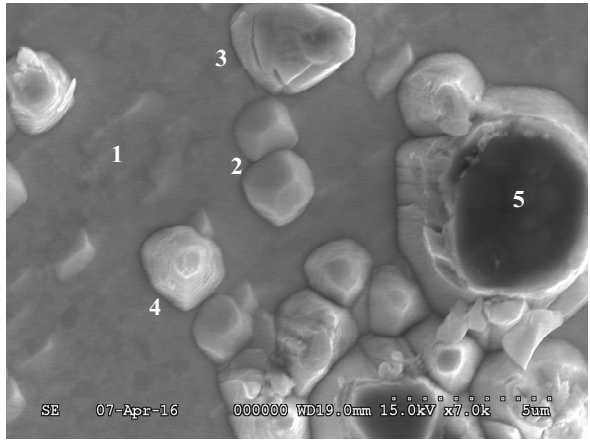

**Fig. 6.** SEM image of the fragment of lateral surface of PbTe crystal after sputtering by $Ar^+$ plasma with energy of 50 eV during 50 min investigated by EDX method.

must be suppressed, and this is what we are observing.

The composition of the re-deposited microscopic structures that have overcome the potential barrier of the critical nucleus formation must be changing during their growth, reflecting the changes of the original surface due to re-sputtering. To evaluate these changes, we carried out EDX analysis of the sample subjected to sputtering for 50 minutes by Ar plasma with energy of 50 eV. Under these sputtering conditions integrated Te sputtering yield was higher by a factor of six compared to the integrated sputtering yield of Pb [12]. Therefore, one might expect measurable changes in composition of sputtered surface. The surface composition at ten statistically random points of both the un-sputtered and sputtered surfaces, as well as the composition of ten surface structures associated with the dimple relief of sputtered surface, was estimated. The view of studied sputtered surface is shown in Fig. 6.

As might be expected, within the accuracy of the method composition of the un-sputtered surface is stoichiometric. The composition of sputtered surface averaging over ten measurements at points similar to the point 1 (Fig. 6) was found to be slightly enriched with lead, namely $Pb_{0.517}Te_{0.483}$. Such result is also to be expected, given strong Te enrichment of the sputtered phase. With regard to the re-deposited surface structures the situation appeared less definitive. The composition of the predominant pyramidal structures similar to the structures 2 (Fig. 6) was very similar to the composition of sputtered surface, but somewhat closer to stoichiometric ratio – its average value was found to be $Pb_{0.508}Te_{0.492}$. The surface composition of conical structures, similar to the structure 3 (Fig. 6), was found to be strongly enriched with lead – maximum enrichment was estimated as $Pb_{0.63}Te_{0.37}$. On the other hand, the surface structures, similar to the structure 4 (Fig. 6), appeared strongly enriched with tellurium (maximum enrichment $Pb_{0.39}Te_{0.61}$). Dark flat surface similar to the point 5 (Fig 6) of the re-deposited structures without any defined shape and spatial orientation contained a lot of Si, O, and C.

## Conclusions

Morphology of lateral surface of the PbTe crystal grown from melt by the Bridgman method and sputtered by Ar plasma with ion energy 50-550 eV under SNMS conditions for 5-50 minutes is investigated. The sputtering PbTe crystal surface in the process of depth profiling was found to be simultaneously both the source of sputtered material and the efficient substrate for re-deposition of the sputtered material. Consequently the sputtered phase of PbTe crystal to be analyzed by SNMS method is the result of superposition of two interdependent processes, namely sputtering and re-deposition.

Sputtering of PbTe crystal forms the dimple relief on its surface. Re-deposition of sputtered Pb and Te leads to covering of the sputtering surfaces with an array of small microscopic surface structures, like hillocks, pyramids, cones and others for any duration of PbTe crystal sputtering by Ar plasma of any energy. The surface structures originate on both the natural growth defects and dimples of sputtered lateral surfaces of the PbTe crystal samples. Under constant sputtering energy, the smaller is the duration of surface sputtering, the smaller is the average size of re-deposited microscopic surface structures, but the higher is their surface density. The same behaviour of the re-deposited microscopic surface structures take place under constant sputtering time of PbTe samples when the sputtering energy changes.